\documentclass[traditabstract]{aa}
\usepackage{ulem}
\usepackage{graphicx}
\usepackage{txfonts}
\usepackage{natbib}
\usepackage{verbatim}
\usepackage{url}
\bibpunct{(}{)}{;}{a}{}{,}
\newcommand{\teff}{$T_{\rm eff}$}
\newcommand{\logg}{$\log g$}
\newcommand{\feh}{[\rm{Fe}/\rm{H}]}
\newcommand{\Vmic}{$\xi_{\rm t}$}

\newcommand{\mni}{Mn\,{\textsc i}\rm}

\newcommand{\mnii}{Mn\,{\textsc {ii}}\rm}
\newcommand{\mnfe}{\rm [Mn/Fe]}

\newcommand{\Mch}{$\rm M_{ch}$}

\newcommand{\Elow}{E_{\rm low}}

\begin{document}

   \title{Observational constraints on the origin of the elements}

   \subtitle{III. Evidence for the dominant role of sub-Chandrasekhar SN Ia in the chemical evolution of Mn and Fe in the Galaxy}

   \author{P.~Eitner
          \inst{1,2}
          \and
          M.~Bergemann\inst{1}
          \and 
          C.~J.~Hansen \inst{1}
          \and 
          G.~Cescutti \inst{3,4}
          \and
          I.~R.~Seitenzahl \inst{5}
          \and 
          S.~Larsen \inst{6}
          \and
          B.~Plez \inst{7}}

   \institute{
Max-Planck Institute for Astronomy, 69117, Heidelberg, Germany\\
\label{inst1}
\and
Ruprecht-Karls-Universität, Grabengasse 1, 69117 Heidelberg, Germany\\
\label{inst2}
\and 
INAF, Osservatorio Astronomico di Trieste, Via G.B. Tiepolo 11, I-34143 Trieste, Italy\\ 
\label{inst3}
\and
IFPU - Institute for Fundamental Physics of the Universe, Via Beirut 2, 34014 Trieste, Italy\\
\label{inst4}
\and
The University of New South Wales, School of Science, Australian Defence Force Academy, Northcott Drive, Canberra 2600, Australia\\
\label{inst5}
\and
Department of Astrophysics/IMAPP, Radboud University, Heyendaalseweg 135, 6525 AJ Nijmegen, The Netherlands\\
\label{inst6}
\and
LUPM, UMR 5299, Universit\'e de Montpellier, CNRS, 34095 Montpellier, France\\
\label{inst7}
}

\date{Received date; accepted date}

 
  \abstract
   {The abundance ratios of manganese to iron in late-type stars across a wide metallicity range place tight constraints on the astrophysical production sites of Fe-group elements. In this work, we investigate the chemical evolution of Mn in the Milky Way galaxy using high-resolution spectroscopic observations of  stars in the Galactic disc and halo stars, as well as a sample of globular clusters. Our analysis shows that local thermodynamic equilibrium (LTE) leads to a strong imbalance in the ionisation equilibrium of Mn I and Mn II lines. Mn I produces systematically (up to 0.6 dex) lower abundances compared to the Mn II lines. Non-local thermodynamic equilibrium (NLTE) radiative transfer satisfies the ionisation equilibrium across the entire metallicity range, $-3 \lesssim \rm{[Fe/H]} \lesssim -1$, leading to consistent abundances from both ionisation stages of the element. We compare the NLTE abundances with Galactic Chemical Evolution models computed using different sources of type Ia and type II supernova (SN Ia and SN II) yields. We find that a good fit to our observations can be obtained by assuming that a significant (${\sim} 75\%$) fraction of SNe~Ia stem from a sub-Chandrasekhar (sub-\Mch) channel. While this fraction is larger than that found in earlier studies (${\sim} 50\%$), we note that we still require ${\sim} 25\%$ near-\Mch\ SNe~Ia to obtain solar [Mn/Fe] at [Fe/H] $= 0$. Our new data also suggest higher SN II Mn yields at low metallicity than typically assumed in the literature.}
  %
  %


   \maketitle
%
%
\section{Introduction}
Manganese is one of the key Fe-group elements that have been extensively studied in the astronomical literature. It has only one stable isotope, $^{55}$Mn, which is neutron rich (odd-Z), and its cosmic production is thought to be predominantly associated with type Ia supernovae (SNe~Ia). Therefore, the evolution of Mn relative to Fe in the Galactic stars is a powerful probe of the epoch when SNe~Ia started contributing to the chemical enrichment and, therefore, of star formation history of the Galactic populations \citep[e.g.][]{Nomoto2013,McWilliam2016,Barbuy2018}.

For SNe~Ia, there are two main channels discussed in the literature. The first is a  single-degenerate (SD) channel  \citep{Whelan1973}, in which a white dwarf (WD) accretes material from another star in a binary system and approaches the Chandrasekhar mass (hereafter, near-\Mch). The second, so-called sub/super-\Mch\ channel involves two possible scenarios \citep{Iben1984, Maoz2014,Levanon2015,Rebassa2019}. In a close binary system with an initial separation of a few Earth radii, a less massive WD can merge with another WD due to the emission of gravitational waves following a common envelope phase, producing a violent merger in a double-degenerate (DD) system \citep{Iben1984, Pakmor2012}. Alternatively, the WD obtains He from the companion that eventually triggers a double detonation, first in the outer He layer and subsequently in the CO core, which liberates enough energy to unbind the star; this system can be either SD or DD \citep[e.g.][]{Livne1990,Fink2010,Shen2018,Goldstein2018}. Also, a scenario in which two WDs collide due to the Lidov-Kozai mechanism has been proposed \citep{Katz2012, Kushnir2013}, although the frequency of the collisions is still debated \citep{toonen2018}.

There are arguments \citep[e.g.][]{Ruiter2009,Gilfanov2010, Shappee2013, Goldstein2018, Kuuttila2019, Flors2019} supporting the idea that a large fraction of SNe~Ia follow from a sub-\Mch\ channel. In this context, both double detonation through He-mass transfer (as SD or DD) and violent merger DD channels are considered as viable. On the other hand, recent X-ray observations of the hot intra-cluster medium \citep[e.g.][]{Mernier2016,Hitomi2017} and SN~Ia light curve properties \citep{Scalzo2014, Scalzo2019} call for a sizeable contribution from near-\Mch\ SNe~Ia.

Despite major differences in the explosion physics and progenitor scenarios, different SN Ia channels generally produce relatively similar chemical abundance distributions in explosive nucleosynthesis, with a few exceptions, such as Mn  \citep{Seitenzahl2013}. Mn is produced in enhanced, that is, super-solar [Mn/Fe] amounts in normal freeze-out from nuclear statistical equilibrium (NSE), where low entropy prevents the conversion of its parent nucleus $^{55}$Co by (p,$\gamma$)-reactions to $^{56}$Ni \citep[]{Jordan2003}. The physical conditions for this low-entropy freeze-out regime require high nuclear fuel densities, in excess of $2 \times 10^8$ g cm$^{-3}$ \citep{Thielemann1986}. From the diverse list of explosion models for normal SNe Ia, only near-\Mch\ SD explosion models have sufficiently high fuel densities to allow large production of Mn \citep[e.g.][]{Hillebrandt2013, Seitenzahl2017}. The high Mn yield of near-\Mch\ SNe~Ia coupled with core-collapse (CC) type II SNe yields for Fe and Mn may therefore produce a characteristic upturn in the chemical evolution trend of [Mn/Fe] versus [Fe/H] in stellar populations of the Milky Way or in other galaxies. On the other hand, the absence of such an upturn may signify the importance of sub- and super-\Mch\ systems in  Galactic Chemical Evolution (GCE). This picture is generally accepted but only qualitatively correct, as it depends, among other parameters, such as the detailed star formation history of the halo and disc, on the details of SN explosion models. For example, the higher metallicity sub-\Mch\ model of \citet{Shen2018} produces super-solar [Mn/Fe] at the low-mass end ($M < 0.9 \mathrm{M}_{\odot}$) of the WD mass distribution, although these events would be too faint to explain spectroscopically normal SNe~Ia, which are expected to constitute the dominant SN~Ia channel. Nevertheless, this qualitative argument has commonly been employed in the literature to explore the contribution of different SN~Ia types to enrichment of Fe-peak elements in different stellar populations  \citep[e.g.][]{Kirby2019, Kobayashi2019, Mia2020}.

In recent years, several studies have investigated [Mn/Fe] abundance ratios in the Galactic stars \citep[e.g.][]{Spite2013, Battistini2015,Mishenina2015}. Local Thermodynamic Equilibrium (LTE) models suggest that the [Mn/Fe] ratio is highly sub-solar in low-metallicity stars \citep{Bonifacio2009,Mishenina2015}, dropping to [Mn/Fe]$\sim -0.6$ dex at [Fe/H] $\sim -2$. However, evidence for departures from LTE stems from the strong excitation imbalance \citep{Cayrel2004, Bergemann2008, Spite2013}, as well as from a significant systematic difference between the Mn abundances measured in turn-off stars and in red giants in the same Galactic population \citep{Bonifacio2009}. Detailed NLTE studies show that \mni\ lines are severely affected by NLTE effects. In particular, \cite{Bergemann2008} showed that a strong depletion of [Mn/Fe] in metal-poor stars is a bias caused by assumption of LTE, and the NLTE trend of [Mn/Fe] is close to solar in the [Fe/H] range from $-2.5$ to $0$.

This is the third paper in our series of studies \citep{Bergemann2019, Gallagher2019}, in which we aim to provide robust observational constraints on the origins of elements and their detailed chemical evolution in the Galaxy. In this paper, we use our new methods to determine Mn abundances in Galactic stars across a wide metallicity range, $-4 \lesssim \rm{[Fe/H]} \lesssim 0$, in order to place constraints on SN Ia progenitors and their explosion mechanism. We follow the methods developed in (\citealt{Bergemann2019}; hereafter, Paper 1). The structure of the paper is as follows. In Sect. \ref{sec:observations}, we describe the observed spectra and their reduction. Section \ref{sec:methods} outlines the methods of the abundance analysis and the main results are summarised in Sect. \ref{sec:results}. In particular, we discuss our findings in the context of nucleosynthesis of Mn, its chemical evolution in the Galaxy, and constraints on the progenitors and explosion mechanism of SN Ia systems.
\section{Observations} \label{sec:observations}
\subsection{Main stellar sample}
The goals of our study place certain requirements on the properties of the observed sample of stars. Firstly, understanding the chemical evolution of the Galaxy requires a broad metallicity coverage to probe the halo and disc regimes. Secondly, systematic uncertainties, which are the dominant source of error in abundance analyses, must be understood. With one caveat (see section 3.2), the ionisation balance method is a powerful technique to test for such biases. The abundances derived from the lines of neutral species are compared with those obtained from the lines of singly ionised species. Once the surface gravity of a star is fixed using a  model-independent method, such as astrometry, the consistency of abundances derived from two ionisation stages represents a powerful diagnostic of the accuracy of abundances. In the case of Mn, the diagnostic lines of \mni\ are located in the optical wavelength range, whereas useful \mnii\ features are found in the near-UV, such as the 3488\,\AA\ line. Therefore, we need spectra covering the entire wavelength window from the near-UV to optical. Owing to extreme line blending in the near-UV, high-resolution, high signal-to-noise ratio (S$/$N) spectra are needed to allow a careful de-blending of the diagnostic absorption features. In addition, stellar parameters must be determined very accurately, employing consistent methods, and validated on detailed NLTE calculations.

\begin{table*}[!ht]
\begin{center}
\setlength{\tabcolsep}{0.016\linewidth}
\caption{Sample of stars with their physical parameters, including LTE and NLTE abundances of Mn. The NLTE abundances are averaged over \mni\ and \mnii\ lines, because of the weak ionisation imbalance. LTE abundances are given either for \mni\ or \mnii\ lines (marked with an asterisk), depending on the available data. The lines of the 4030\AA\ triplet are excluded for giants, since they are sensitive to convection effects (see Paper 1).}
\label{tab:sample}
\begin{tabular}{l c c c c  c c  c}
\hline
\noalign{\smallskip}
star & $\rm T_{eff}$ & $\log g$ & $\rm [Fe/H]$ & vmic    & \multicolumn{2}{c}{$\rm [Mn/Fe]$} \\ 
     &   K   &  dex     &  &  km/s & LTE & NLTE & source \\
\noalign{\smallskip}
\hline
\noalign{\smallskip}

Dwarfs &  & & & & & & \\
HD 3567    &  6035  &  4.08 &  -1.29 &  1.5 & -0.43  & -0.29 & c \\
HD 6582    &  5387  &  4.45 &  -0.83 &  0.9 & -0.21  & -0.13 & a \\ 
HD 19445   &  5982  &  4.38 &  -2.10 &  1.4 & -0.40  & -0.21 & c \\
HD 22879   &  5792  &  4.30 &  -0.95 &  1.2 & -0.24  & -0.15 & c \\ 
HD 84937   &  6350  &  4.10 &  -2.15 &  1.4 & -0.34  & -0.16 & b \\
HD 106038  &  5950  &  4.33 &  -1.45 &  1.1 & -0.22$^*$/-0.4 & -0.15 & c \\
HD 121004  &  5711  &  4.46 &  -0.71 &  0.7 & 0.28$^*$ & 0.23 & c \\
HD 122196  &  6048  &  3.89 &  -1.75 &  1.2 & -0.45  & -0.28 & c \\
BD 133442  &  6450  &  4.42 &  -2.47 & 1.5 & -0.38$^*$ & -0.32 & c \\
HD 134169  &  5930  &  3.98 &  -0.86 &  1.8 & -0.21  & -0.10 & a \\
HD 140283  &  5777  &  3.70 &  -2.38 &  1.3 & -0.51  & -0.21 & b,c \\
HD 142267  &  5807  &  4.42 &  -0.46 &  1.0 & -0.17  & -0.10 & a \\       
HD 144061  &  5815  &  4.44 &  -0.31 &  1.2 & -0.01  & ~0.04 & a \\         
HD 148816  &  5880  &  4.07 &  -0.78 &  1.2 & -0.32  & -0.22 & a \\
HD 157466  &  5990  &  4.38 &  -0.44 &  1.1 & -0.23  & -0.15 & a \\       
HD 158226  &  5805  &  4.12 &  -0.56 &  1.1 & -0.18  & -0.09 & a \\      
HD 160693  &  5850  &  4.31 &  -0.60 &  1.2 & -0.18  & -0.09 & a \\       
HD 160933  &  5765  &  3.85 &  -0.27 &  1.2 & -0.24  & -0.15 & a \\      
HD 170357  &  5665  &  4.07 &  -0.50 &  1.2 & -0.24  & -0.14 & a \\		    
HD 171620  &  6115  &  4.20 &  -0.50 &  1.4 & -0.17  & -0.08 & a \\  
HD 182807  &  6100  &  4.21 &  -0.33 &  1.4 & -0.13  & -0.05 & a \\       
HD 184448  &  5765  &  4.16 &  -0.43 &  1.2 & ~0.01  & ~0.07 & a,c \\      
HD 186379  &  5865  &  3.93 &  -0.41 &  1.2 & -0.20  & -0.10 & a \\    
HD 198300  &  5890  &  4.31 &  -0.60 &  1.2 & -0.04  & ~0.04 & a \\      
HD 200580  &  5940  &  3.96 &  -0.82 &  1.4 & ~0.09  & ~0.19 & a \\      
HD 204155  &  5815  &  4.09 &  -0.66 &  1.2 & -0.34  & -0.21 & a \\
HD 208906  &  6025  &  4.37 &  -0.76 &  1.4 & -0.22  & -0.11 & a \\       
HD 215257  &  6030  &  4.28 &  -0.58 &  1.4 & -0.26  & -0.15 & a \\   	
HD 218209  &  5665  &  4.40 &  -0.60 &  1.1 & -0.05  & ~0.04 & a \\	   
HD 221876  &  5865  &  4.29 &  -0.60 &  1.2 & -0.32  & -0.22 & a \\  
HD 224930  &  5480  &  4.45 &  -0.66 &  0.9 & -0.33  & -0.24 & a \\	
G 64-12    &  6464  &  4.30 &  -3.12 &  1.5 & -0.40$^*$ & -0.32 & b,c \\
Giants &  & & & & & & \\
HD 74462   &  4590  &  1.98 &  -1.43 &  1.1 & -0.36  & -0.07 & c \\
HD 115444  &  4785  &  1.71 &  -2.87 &  2.1 & -0.75  & -0.36 & d \\
HD 122563  &  4665  &  1.65 &  -2.50 &  1.8 & -0.71  & -0.22 & b,c \\
HD 126238  &  4900  &  2.02 &  -1.85 &  1.5 & -0.45  & -0.14 & c \\
HD 126587  &  4950  &  2.36 &  -2.86 &  1.7 & -0.70  & -0.27 & c \\
HD 128279  &  5200  &  1.73 &  -2.13 &  1.3 & -0.55$^*$ & -0.52 & d \\
HD 175305  &  5100  &  2.70 &  -1.34 &  1.2 & -0.23  & ~0.02 & c \\
HD 186478  &  4730  &  1.56 &  -2.33 &  1.8 & -0.53  & -0.21 & d \\
BD +541323 &  5213  &  2.20 &  -1.55 &  1.5 & -0.55  & -0.25 & d \\
HE 0315+0000 & 5050 &  2.47 &  -2.67 &  1.6 & -0.55  & -0.11 & c \\
\noalign{\smallskip}
\hline
\multicolumn{8}{l}{\footnotesize{a: \cite{Gehren2004}, b: \cite{Bergemann2012}, c: \cite{Hansen2013}, d: \cite{Hansen2012,Hansen2011}}}
\end{tabular}
\end{center}
\end{table*}

With these considerations in mind, we compiled our stellar sample from several sources, including \citet{Gehren2004}, \citet{Bergemann2012}, \citet{Hansen2011},  \citet{Hansen2012}, and \citet{Hansen2013}, to cover a broad range in metallicity, from $-4$ to solar (Table \ref{tab:sample}). The sample comprises 42 stars. Only 10 of them are red giants, but most stars are FG dwarfs. The spectra of 10 stars in the sample were taken with the UVES spectrograph on the ESO Very Large Telescope (VLT,  \citealt{Dekker2000}) between 2000 and 2002 with high- resolution $R>40\,000$ and high-S$/$N $>100$ per pixel at 3200\,\AA. These spectra cover a broad wavelength range from 3050 to 6800\,\AA\ with a few gaps owing to the various settings of the spectrograph. Another 8 stars were observed at high resolution with HIRES/Keck. For further details of these spectra we refer to \citet{Hansen2011} and \citet{Hansen2012,Hansen2013}. For the remaining 24 stars, we used the high-quality FOCES data \citep[see e.g.][for more details ]{Gehren2004}. These spectra have a resolving power of $\sim 60\,000$ and a typical S$/$N $> 200$ at 6000\,\AA\ \citep{Gehren2004, Gehren2006}. The spectra of HD 84937, HD 122563, and HD 140283 were taken from the UVES-POP database \citep{Bagnulo2003}.

Stellar parameters of the chosen stars were adopted from the aforementioned studies. The \teff~and \logg\ rely on photometry and astrometry. Metallicity and micro-turbulence values were computed using NLTE analysis of Fe I and Fe II lines \citep{Bergemann2012,Hansen2013}. The metallicities of stars adopted from the sample of \citet{Gehren2004, Gehren2006} rely on the LTE analysis of Fe II lines, but LTE is a reliable assumption for the lines of singly ionised iron \citep[e.g.][]{Bergemann2012,Lind2012,Lind2017}. Finally, four stars (HD 115444, HD 128279, HD 186478, BD$+151323$) have metallicities based on LTE Fe abundances. We include them in the LTE analysis of [Mn/Fe], but do not use them in the NLTE calculations.
\subsection{Additional datasets}
To complement our sample at low metallicity, we add 19 metal-poor main-sequence stars from \citet{Bonifacio2009}, who provide metallicities and Mn abundances measured assuming LTE. However, we correct this sample for NLTE, using the abundance corrections determined as described in Section 3.1. 

In addition, we include [Mn/Fe] abundances for seven Galactic globular clusters, obtained from integrated-light observations with UVES at VLT \citep{Larsen2017} and 15 extragalactic GCs observed with UVES and with HIRES on the Keck I telescope \citep{Larsen2018}. These abundance measurements were made by fitting custom-computed simple stellar population models to the integrated-light spectra, adjusting the input abundances in the models until the best fits to the data were obtained. The synthetic spectra used in these latter models were computed with the ATLAS9 and SYNTHE codes \citep{Sbordone2004, Kurucz2005}. Abundances for Mn were measured from fits to two spectral windows, from 4750 to 4790 \AA\ and 6010 to 6030 \AA. For further details we refer to \citet{Larsen2017}. The NLTE corrections for the integrated light were computed as described in our recent study \citep{Eitner2019}.

\section{Methods} \label{sec:methods}
\subsection{Determination of Mn abundances}
The Mn abundances are computed as follows. For most stars in the sample, we first use the MOOG code \citep[][version 2014]{Sneden2016} to obtain LTE abundances via spectrum synthesis and then we correct these estimates for NLTE effects. MARCS model atmospheres \citep{Gustafsson2008} are employed. The MOOG code is capable of synthesising larger wavelength regions, which is needed in crowded regions, where line blends may occur (especially in the more metal-rich stars). This allows us to take all known blends into consideration, as well as to check the local placement of the continuum, which is difficult in the blue wavelength region. The list of diagnostic Mn lines is provided in Table \ref{tab:lines}. All atomic data were taken from Paper 1. We note that owing to the limited spectral coverage and gaps in the observed spectra, only a subset of these lines were used for every star.
\begin{table}
\begin{minipage}{\linewidth}
\renewcommand{\footnoterule}{} 
\caption{Parameters of lines used for abundance calculation. The atomic data are taken from Paper 1.}
 \label{tab:lines}     
\begin{center}
\begin{tabular}{lrr}
\noalign{\smallskip}\hline\noalign{\smallskip}  ~~~~~~$\lambda$ & $\Elow$ & $\log gf$ \\
     ~~~~[\AA] &  [eV]  &   \\
\noalign{\smallskip}\hline\noalign{\smallskip}
  3488.68 &  1.85   &  --0.937 \\
  3496.81 &  1.83   &  --1.779 \\
  3497.53 &  1.85   &  --1.418 \\
  4030.76 &  0.00   &  --0.497 \\
  4033.07 &  0.00   &  --0.647 \\
  4034.49 &  0.00   &  --0.843 \\
  4055.54 &  2.14   &  --0.077 \\
  4070.28 &  2.19   &  --1.039 \\
  4451.58 &  2.89   &    0.278 \\
  5394.67 &  0.00   &  --3.503 \\  
  5407.42 &  2.13   &  --1.743 \\  
  5420.35 &  2.13   &  --1.462 \\  
  5432.54 &  0.00   &  --3.795 \\  
  6013.49 &  3.06   &  --0.251 \\  
  6016.64 &  3.06   &  --0.216 \\  
  6021.79 &  3.06   &    0.034 \\  
\noalign{\smallskip}\hline\noalign{\smallskip}
\end{tabular}
\end{center}
\end{minipage}
\end{table}

The NLTE corrections to abundances, $\Delta_{NLTE} = A_{NLTE} - A_{LTE}$, were computed with MULTI2.3 code \citep{Carlsson1986} as described in \citet{Eitner2019}. The LTE and NLTE grids were constructed for a grid of different Mn abundances for each star in the sample. By calculating the equivalent widths (EW) of the lines and applying linear interpolation, we obtained theoretical curves-of-growth as a function of Mn abundance. We then employed the measured LTE abundance values to find the closest matching NLTE EW for each Mn I line and to extract the corresponding NLTE abundances.

For four metal-poor stars, which are not affected by blending, we used the measured EWs of Mn lines directly to infer the LTE and NLTE abundances using the LTE and NLTE grids computed as described above. The EWs were measured in IRAF by fitting Gaussian functions to the Mn lines. The integration and interpolation procedures were taken from the \texttt{scipy} library \citep{Oliphant2007}. 

In both cases, we used the same model atmospheres and atomic data to ensure consistency in the calculations. We  also verified that both procedures return the same abundance values for a spectrum, which is not affected by blending.
\subsection{Systematic and statistical uncertainties}
There are several sources of abundance uncertainties \citep[e.g.][]{Bergemann2012}. Systematic uncertainties are caused by  approximations used to model physics of stellar atmospheres and radiation transfer. These have an impact on the estimates of stellar parameters and abundances. Statistical uncertainties usually reflect the imperfections of the observed data, which stem from instrumental effects and data reduction procedures. We also take into account the statistical fluctuations in the EW measurement process. All these effects result into $0.13$ dex statistical uncertainty for our sample \citep[see][]{Hansen2013}. In what follows, we explore the sensitivity of our Mn abundances to the uncertainties in stellar parameters. We also assess the systematic error caused by adopting 1D hydrostatic models and LTE in line formation. To the best of our knowledge, this is the first study of the Galactic chemical evolution of [Mn/Fe], in which both sources of error are treated in detail.
\begin{table}
\label{tab:errors}
\caption{Variation of individual stellar parameters and their impact on various Mn lines' abundances in HD 106038.}
\setlength{\tabcolsep}{0.01\linewidth}
\centering
\begin{tabular}{c c c c c c}
\hline
\noalign{\smallskip}
 Parameter & $\Delta$ & 3488\AA & 3497\AA & 3498\AA & 6021\AA \\ 
\noalign{\smallskip}
\hline
\noalign{\smallskip}
\teff  & $-100$ K &  $-0.04$ & $-0.05$ & $-0.07$ & $-0.10$  \\
$\log$ g  & $-0.2$ dex &  $-0.04$ & $-0.06$ & $-0.03$ & $-0.04$ \\
V$_{\rm mic}$ & $-0.15$ kms$^{-1}$ &  ~~0.10  & ~~0.04  & ~~0.06  & $-0.04$ \\ $\rm{[Fe/H]}$ & $-0.1$ dex &  $-0.02$ & $-0.01$ & $-0.01$ & $-0.03$ \\
\noalign{\smallskip}
\hline
\end{tabular}
\end{table}

A comprehensive analysis of the uncertainties of \teff, \logg, \Vmic, and \feh\ is given in \citet{Gehren2004}, \citet{Gehren2006}, \citet{Bergemann2012}, and \citet{Hansen2013}. We adopt their errors to carry out a systematic analysis of their impact on the Mn abundance estimates. The changes of Mn abundance caused by the variation of stellar parameters within their uncertainty ranges are given in Table \ref{tab:errors}. We have chosen a main-sequence star HD 106038 for this test, as this star is representative of the majority of our targets and it has a relatively low metallicity ([Fe/H] $=-1.45$ dex). Our tests suggest that \mni\ and \mnii\ lines are relatively insensitive to the variation of stellar parameters within their uncertainty margins. For example, the variation of \teff\ by 100 K causes a maximum change of the Mn abundance by $0.1$ dex, whereas the sensitivity of Mn lines to surface gravity and metallicity is even lower than that. The variation of microturbulence by $\pm 0.15$ kms$^{-1}$ is only significant for the \mnii\ 3488 \AA\ line, which yields a $0.1$ dex higher or lower abundance.

Sensitivities of Mn abundance measurements to the uncertainties of stellar parameters are also discussed in \citet[][Tables 6,7]{Cayrel2004} and in \citet[][Table 4]{Bonifacio2009}. The results of these latter authors are very similar to our estimates for HD 106038, although their estimates refer to very metal-poor giants and dwarfs with metallicity [Fe/H] $\sim -3$. Combining the errors in quadrature, we estimate the uncertainty of Mn abundance caused by the uncertainty of stellar parameters to be 0.06 dex for Mn I lines and 0.05 dex for Mn II lines. One of our objects, HD 22879, was recently analysed by \citet{Mishenina2015}. Their stellar parameters are somewhat different from our values. We therefore  re-computed the $\mnfe$ ratios for HD 22879 using two different model atmospheres: one (our study) with \teff $= 5792$ K, \logg $= 4.30$ dex, \feh $= -0.95$, and \Vmic\ $= 1.2$~km/s. For the other model, we assumed \teff $= 5792$ K, \logg\ $= 4.50$ dex, \feh $= -0.77$, and \Vmic\ $= 1.1$ km/s, as recommended by \citet{Mishenina2015}. We found that the Mn abundances derived using both sets of parameters are consistent to $0.02$ dex. This confirms that our results are robust to the uncertainties in stellar parameters.

A detailed discussion of the differences between 1D LTE, 1D NLTE, 3D LTE, and 3D NLTE abundances of Mn can be found in Paper 1. Generally, the differences between 1D NLTE and 3D NLTE Mn abundances are not large and do not exceed $0.1$ dex for dwarfs and giants with $\feh \geq -2$ (see Fig. 18 in Paper 1). However, some \mni\ lines, in particular the 403 nm resonance triplet, are significantly affected by convection, especially at low metallicity. Is not yet possible for us to perform 3D NLTE calculations for a large stellar sample, as such calculations are very computationally costly. Hence, in order to reduce the influence of convection as far as possible, we exclude the 403 nm triplet lines from the abundance analysis. Abundances derived from the higher-excitation Mn lines are subject to a minor systematic bias only. However, this systematic bias is larger for stars with $\log g \lesssim 2$ dex and $\feh \lesssim -2$ dex. In the extreme case of HD 122563, which has a surface gravity of $1.4$ dex, the bias amounts to $-0.5$ dex. In addition, we showed that for very metal-poor RGB stars both ionisation stages are very sensitive to 3D NLTE, meaning that even a seemingly satisfactory ionisation balance is not a guarantee of unbiased abundances. However, our present sample contains only five such very metal-poor RGB stars (Table \ref{tab:sample}). For them, our 1D NLTE Mn estimates are underestimated and are expected to be closer to the scaled solar value. Our conclusions do not change regardless of whether these stars are included in the sample or not, as the Galactic chemical evolution trend is set by the bulk population of main-sequence stars in our sample.

Our final estimates of uncertainties were computed by summing up the errors caused by the uncertainties of stellar parameters and the statistical errors of the abundance measurements in quadrature. In addition, we account for the systematic errors caused by using imperfect models (1D LTE) or neglecting effects of 3D convection. The systematic error component is strictly positive, as 3D NLTE and 1D NLTE abundances are always larger than 1D LTE abundance. We therefore provide two estimates of uncertainties for each [Mn/Fe] value that correspond to the lower and upper error estimates. The systematic error component is added only to the upper error estimate. As discussed above, we adopt a systematic error of 0.1 dex for our 1D NLTE measurements, whereas the bias caused by 1D LTE amounts to 0.25 dex for dwarfs ($\log g > 3.5$) and 0.4 dex for giants ($\log g < 3.5$). These estimates correspond to our detailed estimates of 3D NLTE and 1D NLTE abundance corrections in Paper 1 (in particular, Fig. 18). For the very metal-poor giants with [Fe/H] $< -2$ dex, we assume a conservative systematic error due to the lack of 3D convection of $0.2$ dex. This yields the maximum upper error estimate of 0.17 dex for 1D NLTE abundances, 0.28 dex for 1D LTE abundances measured in the spectra of dwarfs ($\log g > 3.5$), and 0.42 dex for those of giants ($\log g < 3.5$).
%
\section{Results} \label{sec:results}
\begin{figure}[!tb]
\centering
\caption{Differences between abundances derived from \mni\ and \mnii\ line for the metal-poor Galactic stars, for which UV spectra are available and robust \mnii\ measurements could be derived.}
              \label{fig:ion_eq}%
   \includegraphics[width=0.5\textwidth]{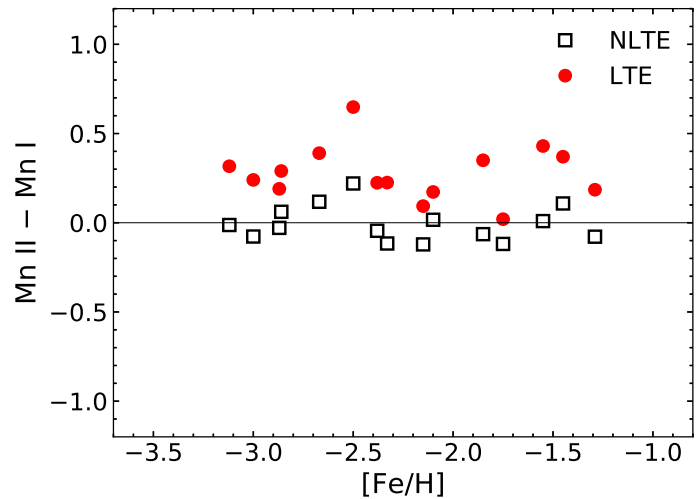}

\end{figure}

Our LTE and NLTE Mn abundances for every star in the main sample are listed in Table~\ref{tab:sample}. Figure  \ref{fig:ion_eq} compares the differences between the abundances derived from the \mni\ and \mnii\ lines. These are available in the UVES spectra only, and therefore only 15 stars are shown. LTE modelling reveals a strong ionisation imbalance, with abundances derived from \mni\ lines being significantly lower compared to the those obtained from \mnii\ lines: the systematic difference ranges from $0.1$ dex to $0.45$ dex. On the other hand, the agreement of two ionisation stages in NLTE is better than $0.1$ dex. This confirms our previous findings in \citet{Bergemann2019}. One-dimensional NLTE modelling leads to significantly increased abundances derived from Mn I lines, which largely solves the problem of ionisation imbalance.

\begin{figure}[!bt]
\centering
\caption{Manganese abundance ratios as a function of metallicity. 
Top panel: LTE data are presented with different symbols;  the line shows the results of the GCE model with yield for SNe II taken from \citet{WW95}, including only near-\Mch\ SNe Ia  with by \citet{Iwamoto99}. 
Bottom panel: NLTE data. The GCE model presented here adopts the yields by \citet{Kobayashi11} for SNe II and includes 25\% of near-\Mch\ and 75\% of sub-\Mch\ SNe Ia.}
\label{fig:Mn_Fe}%
\includegraphics[width=0.49\textwidth]{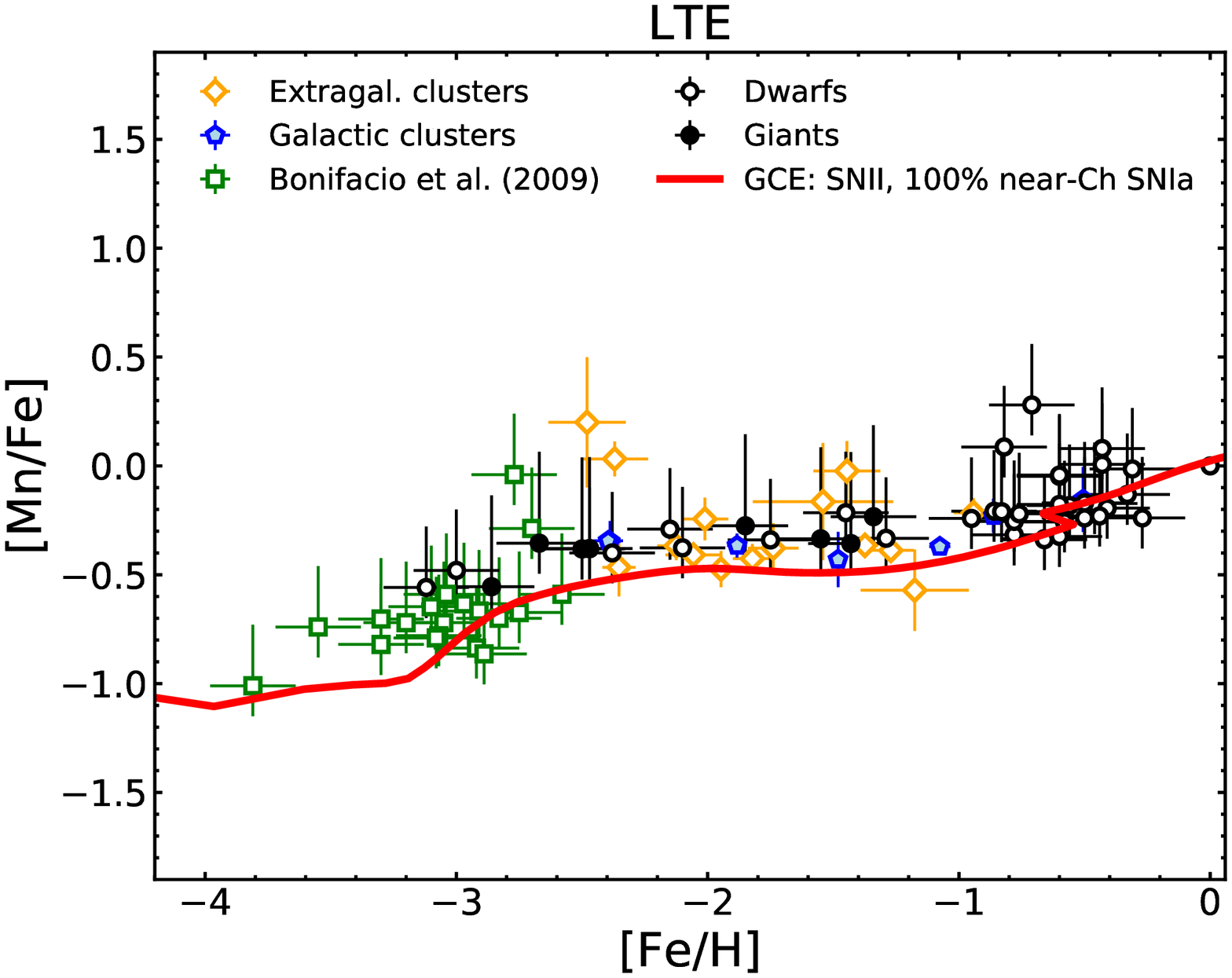}
\includegraphics[width=0.49\textwidth]{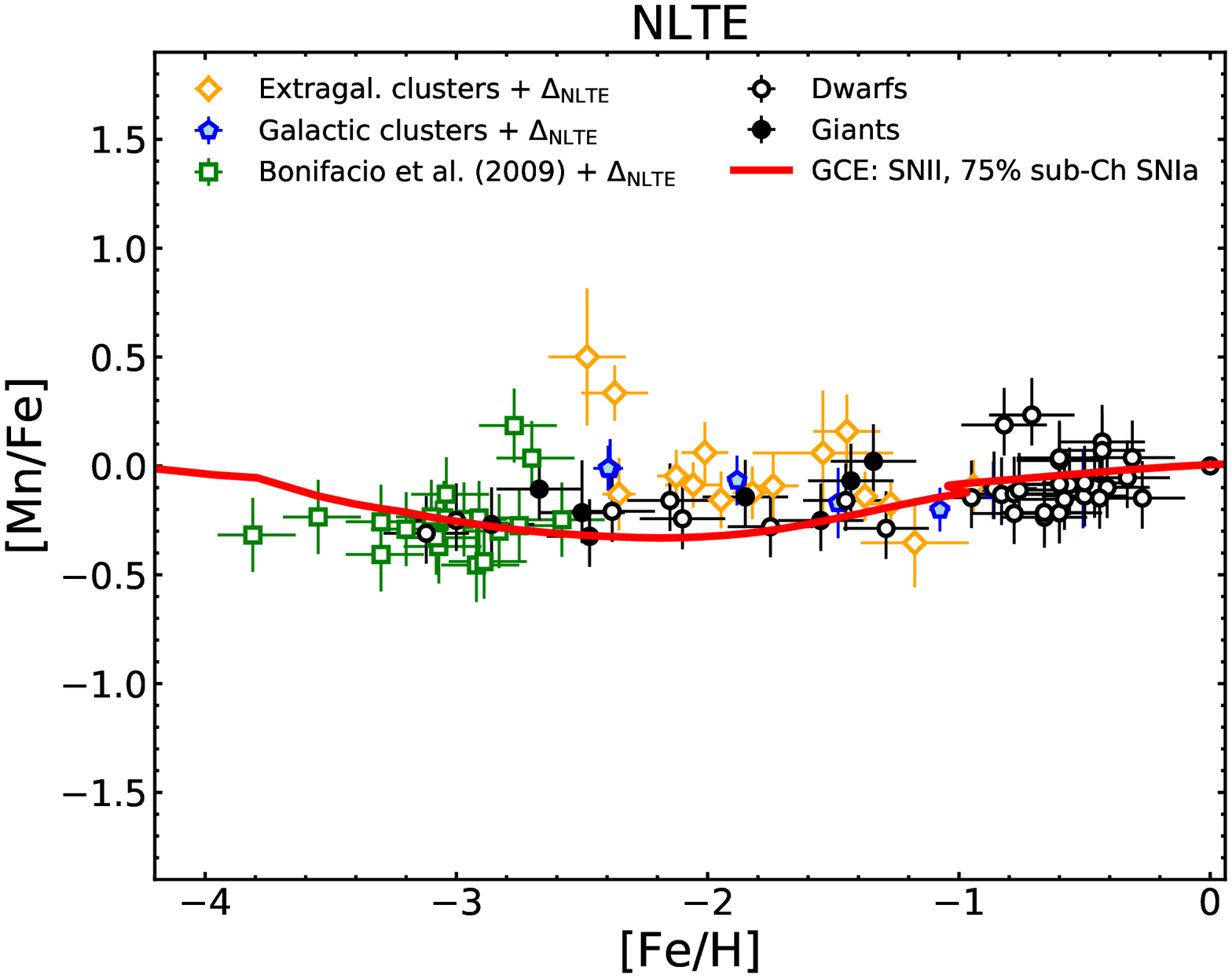}
\end{figure}

Figure \ref{fig:Mn_Fe} shows the measured abundances as a function of metallicity. The abundances in NLTE are averaged over \mni\ and \mnii\ lines. We also include the sample of very metal-poor stars from \citet{Bonifacio2009}  in order to extend our data to very low metallicities. The Mn results from \citet{Bonifacio2009} show the typical signature of the LTE bias, which we correct for NLTE effects using our new Mn model. Our results generally corroborate the findings by \citet{Bergemann2008}: in NLTE, the overall [Mn/Fe] trend remains almost flat and [Mn/Fe] ratios do not deviate significantly from the scaled solar values.

We find a very different picture from our LTE measurements. In line with other LTE studies, we find that the abundances based on the \mni\ lines suggest a strong [Mn/Fe] depletion at low [Fe/H], with abundance ratios dropping to $\sim -0.8$ at [Fe/H] $\sim -3$ and approaching $-1$ dex at [Fe/H] $\sim -4$. On the other hand, we also find that the LTE abundances derived from the \mnii\ lines are very similar to the results we find from these lines in NLTE. In other words, in LTE there is a strong mismatch between the [Mn/Fe] abundance ratios derived from \mni\ and \mnii\ lines. This severe ionisation imbalance, which increases with decreasing metallicity and \logg, generally renders 1D LTE modelling of Mn abundances useless and, therefore, negates the astrophysical analysis of 1D LTE [Mn/Fe] ratios for any accurate study of stellar physics or Galactic chemical evolution.
\section{Discussion}\label{sec:discussion}
Our revision of Mn abundances implies that \mnfe\ ratios across the full metallicity range are much closer to solar (Fig.~\ref{fig:Mn_Fe}, bottom panel) compared to the results obtained under the un-physical assumption of LTE (Fig.~\ref{fig:Mn_Fe}, top panel). Moreover, we only see a weak dependence of \mnfe\ ratios on metallicity. Even at $\feh \approx -4$ the [Mn/Fe]  abundance ratios do not fall below $-0.5$ dex. Adding the sample of very metal-poor stars from \citet{Bonifacio2009} does not change this conclusion. This sample is in excellent agreement with our measurements when corrected for NLTE. 

The data from Galactic and extragalactic globular clusters (\citealt{Larsen2017,Larsen2018}) also follow the [Mn/Fe] trends seen in data of field stars. This is interesting because globular clusters probe different star formation environments and ex-situ origin has been proposed for a substantial fraction, namely one quarter, of the Galactic GC systems \citep{Pritzl2005,Forbes2010}. In addition, the similarity of the abundances for Galactic and extra-galactic GC systems suggests that the trend of [Mn/Fe] with metallicity is universal, which provides  strong evidence against the diversity of stellar explosion mechanisms contributing to the chemical enrichment of Fe-group elements.

Our results have important implications for understanding the sources of Fe-group elements and GCE. The  nucleosynthesis of Fe-group elements, and therefore the chemical enrichment of Mn, is determined by CC SNe and by SNe Ia. There is a general consensus surrounding the qualitative production of Mn and Fe in both sites \citep[e.g.][]{McWilliam2018}. At metallicities  [Fe/H]$\lesssim -$1, the main sources of these elements in the Milky Way are CC~SNe \citep{Cescutti08,Seitenzahl2013}, whereas SNe~Ia, which have longer delay times, dominate the production of both elements at higher metallicity. Quantitatively, however, the picture is still open for debate \citep[e.g.][]{Travaglio2004,Goldstein2018,Kirby2019}. The explosion mechanism of SNe~II is not fully understood \citep[e.g.][]{Janka2012} and there is no consensus on the progenitors of SNe~Ia \citep[e.g.][]{Hillebrandt2013,Maoz2014}. As a consequence, the detailed evolution of \mnfe\ ratio with metallicity in the Galaxy is still largely unknown.

In the top panel of Fig.~\ref{fig:Mn_Fe}, we compare the LTE abundances to a GCE model computed in \citet{Cescutti08}. This model employs the \citet{WW95} yields for SNe II and \citet{Iwamoto99} yields (W7 model) for SNe~Ia. According to this model, the \mnfe\ ratio is extremely low at [Fe/H]$<-$3, but it gradually increases with [Fe/H] owing to the metallicity-dependent yields of CC SNe. At [Fe/H]$\approx -1$, SNe~Ia start contributing to the chemical enrichment and the [Mn/Fe] ratio quickly approaches the solar value. The same behaviour is seen in the LTE data. It should also be noted that the data lie slightly above the model results. Indeed, \citet{Seitenzahl2013} had to increase the CC SN yield of Mn predicted by the \citet{WW95} models by $25\%$ to match the LTE abundances at [Fe/H]$\lesssim -1$.

Our new NLTE data suggest that the [Mn/Fe] trend is flat and the Mn/Fe abundance ratios are close to the scaled solar value at all metallicities (lower panel of Fig.~\ref{fig:Mn_Fe}). This can be accommodated within the available GCE model by increasing the \citet{WW95} CC SN yields for Mn by $50\%$. On the other hand, a GCE model that adopts the \citet{Kobayashi11} yields for normal CC SNe leads to good agreement with our NLTE data. The differences are possibly connected to the different explosion mechanism imposed in the two studies: the piston mechanism in \citet{WW95} and the thermal bomb mechanism in \citet{Kobayashi11}.

Some studies \citep{Romano10,Kobayashi11} also suggest that a $\sim 50\%$ fraction of hypernovae might be necessary to explain the observed super-solar [Zn/Fe] abundance ratios at very low metallicities, [Fe/H] $\lesssim -2$. However, we abstain from this possibility for two reasons. Firstly, the Zn abundances in metal-poor stars rely on the LTE assumption. The NLTE analysis of Zn line formation by \citet{Takeda2005} suggests that the NLTE abundance corrections for Zn I lines are small and positive, but these authors used Drawin's formula to compute the rates of H collisions and the hydrogenic (Kramer's) approximation to compute the rates of photo-ionisation in the statistical equilibrium equations. Both formulae are known to yield woefully inadequate cross-sections and rates for every element that has been studied with detailed quantum-mechanical methods \citep[e.g.][]{Barklem2016,Belyaev2017}. For Mn, for example, Drawin's formula under- or overestimates the rates by up to seven orders of magnitude \citep{Bergemann2019}. Also, Kramer's opacity formula fails to reproduce the overall photoionisation cross-section and its complex frequency structure \citep[e.g.][]{Bautista2017}. Therefore, it is not clear yet whether Zn abundances can be trusted, especially at low metallicity. Secondly, hypernovae tend to severely underproduce Mn compared to Fe. However, their fraction and Galactic population properties are virtually unknown, and must, in turn, be  calibrated on the observed abundances in metal-poor stars. It is also worth pointing out that the yields of rapidly rotating massive stars exploding as CC SNe are different from those of non-rotating objects \citep[][]{LC18}, but  owing to poorly constrained rotation rates of supernova progenitors, detailed quantitative statements on the effect of those are not yet possible.

The larger production of Mn at low metallicity required to explain the nearly solar-like \mnfe\ ratios at [Fe/H] $\lesssim -1$ also has an impact on the expected contribution from SNe~Ia. Our NLTE data suggest only a mild up-turn in the [Fe/H] - [Mn/Fe] plane above $\feh=-1$  compared to previous literature estimates. As a result, the GCE models computed with purely near-\Mch\ SNe do not fit our NLTE data. It is also not possible to reproduce the trend assuming a mix of 50\% near-\Mch\ and 50\% sub-\Mch\ SNe Ia, which was assumed in the favoured model of \cite{Seitenzahl2013}. Our NLTE data (Fig.\ref{fig:Mn_Fe}) suggest that the optimal GCE model for [Mn/Fe] has to assume only $\sim$ 25\% near-\Mch\ and over 75\% sub-\Mch\ SNe Ia. Only in this case can the flat trend of \mnfe\ with [Fe/H] and the solar abundance of Mn  be reproduced simultaneously.

We also find that although there are a few outliers in the \feh-\mnfe\ plane, the scatter of \mnfe\ ratios at any metallicity is relatively small and does not change with \feh. However, our sample is not large enough to draw conclusions on the real astrophysical spread in the disc or halo. Such a spread has been proposed by \citet{cescutti17} as a consequence of multiple types of SNe Ia.
%
%
\section{Conclusions} \label{sec:conclusion}
In this study we calculated NLTE abundances of Mn for stars covering a broad range of metallicities, from solar to the extremely metal-poor regime of $\feh \sim -4$. We analysed the results with respect to the NLTE effects on the Mn ionisation balance. Our new NLTE abundances derived from \mni\ and \mnii\ lines show good consistency, contrary to LTE abundances from previous studies, providing strong evidence that these inconsistencies are mainly due to the assumption of LTE.

Additionally, we investigated the \mnfe\ trend with metallicity in the Milky Way galaxy, in Galactic and in extragalactic globular clusters. We found that the NLTE Mn/Fe ratios are close to solar across the full metallicity range, $-4 \lesssim$ [Fe/H] $\lesssim 0$. These findings contrast those reported from LTE studies. Also, with our NLTE abundances, no steep rise of [Mn/Fe] is visible above $\feh \sim -1$.

A good fit to our data can be obtained by assuming that a significant fraction of SNe Ia (${\sim} 75\%$) stem from the sub-\Mch\ channel. While this fraction is larger than the ${\sim} 50\%$ found by \cite{Seitenzahl2013}, we note that we still require ${\sim} 25\%$ near-\Mch\ SNe~Ia to obtain solar [Mn/Fe] at [Fe/H] $= 0$. In summary, we suggest that the contribution by SN II and SN Ia to Mn/Fe yields has to be different from what is commonly assumed in the literature. Our data indicate that core-collapse SNe produce significant Mn/Fe at lower metallicity ($\feh \lesssim -1$), whereas at higher metallicity a significant contribution by sub-\Mch\ SNe Ia is necessary to simultaneously reproduce the Mn/Fe ratios in the Sun and in the Galactic disc stars.
\begin{acknowledgements}
IRS was supported by the Australian Research Council grant FT160100028. GC acknowledges financial support from the European Union Horizon 2020 research and innovation programme under the Marie Sklodowska-Curie grant agreement No. 664931.
This work has been partially supported by the EU COST Action CA16117 (ChETEC). BP is partially supported by the CNES, Centre National d’Etudes Spatiales. We acknowledge support by the Collaborative Research centre SFB 881 (projects A5, A10), Heidelberg University, of the Deutsche Forschungsgemeinschaft (DFG, German Research Foundation).
\end{acknowledgements}
\bibliography{references}

\begin{thebibliography}{76}
\expandafter\ifx\csname natexlab\endcsname\relax\def\natexlab#1{#1}\fi

\bibitem[{{Bagnulo} {et~al.}(2003){Bagnulo}, {Jehin}, {Ledoux}, {Cabanac},
  {Melo}, {Gilmozzi}, \& {ESO Paranal Science Operations Team}}]{Bagnulo2003}
{Bagnulo}, S., {Jehin}, E., {Ledoux}, C., {et~al.} 2003, The Messenger, 114, 10

\bibitem[{{Barbuy} {et~al.}(2018){Barbuy}, {Chiappini}, \&
  {Gerhard}}]{Barbuy2018}
{Barbuy}, B., {Chiappini}, C., \& {Gerhard}, O. 2018, ArXiv e-prints
  [\eprint[arXiv]{1805.01142}]

\bibitem[{{Barklem}(2016)}]{Barklem2016}
{Barklem}, P.~S. 2016, \aapr, 24, 9

\bibitem[{{Battistini} \& {Bensby}(2015)}]{Battistini2015}
{Battistini}, C. \& {Bensby}, T. 2015, \aap, 577, A9

\bibitem[{{Bautista} {et~al.}(2017){Bautista}, {Lind}, \&
  {Bergemann}}]{Bautista2017}
{Bautista}, M.~A., {Lind}, K., \& {Bergemann}, M. 2017, \aap, 606, A127

\bibitem[{{Belyaev} \& {Voronov}(2017)}]{Belyaev2017}
{Belyaev}, A.~K. \& {Voronov}, Y.~V. 2017, \aap, 606, A106

\bibitem[{{Bergemann} {et~al.}(2019){Bergemann}, {Gallagher}, {Eitner},
  {Bautista}, {Collet}, {Yakovleva}, {Mayriedl}, {Plez}, {Carlsson},
  {Leenaarts}, {Belyaev}, \& {Hansen}}]{Bergemann2019}
{Bergemann}, M., {Gallagher}, A.~J., {Eitner}, P., {et~al.} 2019, \aap, 631,
  A80

\bibitem[{{Bergemann} \& {Gehren}(2008)}]{Bergemann2008}
{Bergemann}, M. \& {Gehren}, T. 2008, \aap, 492, 823

\bibitem[{{Bergemann} {et~al.}(2012){Bergemann}, {Lind}, {Collet}, {Magic}, \&
  {Asplund}}]{Bergemann2012}
{Bergemann}, M., {Lind}, K., {Collet}, R., {Magic}, Z., \& {Asplund}, M. 2012,
  \mnras, 427, 27

\bibitem[{{Bonifacio} {et~al.}(2009){Bonifacio}, {Spite}, {Cayrel}, {Hill},
  {Spite}, {Fran{\c c}ois}, {Plez}, {Ludwig}, {Caffau}, {Molaro}, {Depagne},
  {Andersen}, {Barbuy}, {Beers}, {Nordstr{\"o}m}, \& {Primas}}]{Bonifacio2009}
{Bonifacio}, P., {Spite}, M., {Cayrel}, R., {et~al.} 2009, \aap, 501, 519

\bibitem[{Carlsson(1986)}]{Carlsson1986}
Carlsson, M. 1986, Upps. Astron. Obs. Rep., No. 33, 2+39+117 pp., 33

\bibitem[{{Cayrel} {et~al.}(2004){Cayrel}, {Depagne}, {Spite}, {Hill}, {Spite},
  {Fran{\c{c}}ois}, {Plez}, {Beers}, {Primas}, {Andersen}, {Barbuy},
  {Bonifacio}, {Molaro}, \& {Nordstr{\"o}m}}]{Cayrel2004}
{Cayrel}, R., {Depagne}, E., {Spite}, M., {et~al.} 2004, \aap, 416, 1117

\bibitem[{{Cescutti} \& {Kobayashi}(2017)}]{cescutti17}
{Cescutti}, G. \& {Kobayashi}, C. 2017, \aap, 607, A23

\bibitem[{{Cescutti} {et~al.}(2008){Cescutti}, {Matteucci}, {Lanfranchi}, \&
  {McWilliam}}]{Cescutti08}
{Cescutti}, G., {Matteucci}, F., {Lanfranchi}, G.~A., \& {McWilliam}, A. 2008,
  \aap, 491, 401

\bibitem[{{de los Reyes} {et~al.}(2020){de los Reyes}, {Kirby}, {Seitenzahl},
  \& {Shen}}]{Mia2020}
{de los Reyes}, M. A.~C., {Kirby}, E.~N., {Seitenzahl}, I.~R., \& {Shen}, K.~J.
  2020, arXiv e-prints, arXiv:2001.01716

\bibitem[{{Dekker} {et~al.}(2000){Dekker}, {D'Odorico}, {Kaufer}, {Delabre}, \&
  {Kotzlowski}}]{Dekker2000}
{Dekker}, H., {D'Odorico}, S., {Kaufer}, A., {Delabre}, B., \& {Kotzlowski}, H.
  2000, in Society of Photo-Optical Instrumentation Engineers (SPIE) Conference
  Series, Vol. 4008, \procspie, ed. M.~{Iye} \& A.~F. {Moorwood}, 534--545

\bibitem[{{Eitner} {et~al.}(2019){Eitner}, {Bergemann}, \&
  {Larsen}}]{Eitner2019}
{Eitner}, P., {Bergemann}, M., \& {Larsen}, S. 2019, \aap, 627, A40

\bibitem[{{Fink} {et~al.}(2010){Fink}, {R{\"o}pke}, {Hillebrandt},
  {Seitenzahl}, {Sim}, \& {Kromer}}]{Fink2010}
{Fink}, M., {R{\"o}pke}, F.~K., {Hillebrandt}, W., {et~al.} 2010, \aap, 514,
  A53

\bibitem[{Fl{\"o}rs {et~al.}(2019)Fl{\"o}rs, Spyromilio, Taubenberger, Blondin,
  Cartier, Leibundgut, Dessart, Dhawan, \& Hillebrandt}]{Flors2019}
Fl{\"o}rs, A., Spyromilio, J., Taubenberger, S., {et~al.} 2019, Monthly Notices
  of the Royal Astronomical Society, 491, 2902

\bibitem[{{Forbes} \& {Bridges}(2010)}]{Forbes2010}
{Forbes}, D.~A. \& {Bridges}, T. 2010, \mnras, 404, 1203

\bibitem[{{Gallagher} {et~al.}(2019){Gallagher}, {Bergemann}, {Collet}, {Plez},
  {Leenaarts}, {Carlsson}, {Yakovleva}, \& {Belyaev}}]{Gallagher2019}
{Gallagher}, A.~J., {Bergemann}, M., {Collet}, R., {et~al.} 2019, arXiv
  e-prints, arXiv:1910.03898

\bibitem[{{Gehren} {et~al.}(2004){Gehren}, {Liang}, {Shi}, {Zhang}, \&
  {Zhao}}]{Gehren2004}
{Gehren}, T., {Liang}, Y.~C., {Shi}, J.~R., {Zhang}, H.~W., \& {Zhao}, G. 2004,
  \aap, 413, 1045

\bibitem[{{Gehren} {et~al.}(2006){Gehren}, {Shi}, {Zhang}, {Zhao}, \&
  {Korn}}]{Gehren2006}
{Gehren}, T., {Shi}, J.~R., {Zhang}, H.~W., {Zhao}, G., \& {Korn}, A.~J. 2006,
  \aap, 451, 1065

\bibitem[{{Gilfanov} \& {Bogd{\'a}n}(2010)}]{Gilfanov2010}
{Gilfanov}, M. \& {Bogd{\'a}n}, {\'A}. 2010, \nat, 463, 924

\bibitem[{{Goldstein} \& {Kasen}(2018)}]{Goldstein2018}
{Goldstein}, D.~A. \& {Kasen}, D. 2018, \apjl, 852, L33

\bibitem[{{Gustafsson} {et~al.}(2008){Gustafsson}, {Edvardsson}, {Eriksson},
  {J{\o}rgensen}, {Nordlund}, \& {Plez}}]{Gustafsson2008}
{Gustafsson}, B., {Edvardsson}, B., {Eriksson}, K., {et~al.} 2008, \aap, 486,
  951

\bibitem[{{Hansen} {et~al.}(2013){Hansen}, {Bergemann}, {Cescutti}, {Fran{\c
  c}ois}, {Arcones}, {Karakas}, {Lind}, \& {Chiappini}}]{Hansen2013}
{Hansen}, C.~J., {Bergemann}, M., {Cescutti}, G., {et~al.} 2013, \aap, 551, A57

\bibitem[{{Hansen} \& {Primas}(2011)}]{Hansen2011}
{Hansen}, C.~J. \& {Primas}, F. 2011, \aap, 525, L5

\bibitem[{{Hansen} {et~al.}(2012){Hansen}, {Primas}, {Hartman}, {Kratz},
  {Wanajo}, {Leibundgut}, {Farouqi}, {Hallmann}, {Christlieb}, \&
  {Nilsson}}]{Hansen2012}
{Hansen}, C.~J., {Primas}, F., {Hartman}, H., {et~al.} 2012, \aap, 545, A31

\bibitem[{{Hillebrandt} {et~al.}(2013){Hillebrandt}, {Kromer}, {R{\"o}pke}, \&
  {Ruiter}}]{Hillebrandt2013}
{Hillebrandt}, W., {Kromer}, M., {R{\"o}pke}, F.~K., \& {Ruiter}, A.~J. 2013,
  Frontiers of Physics, 8, 116

\bibitem[{{Hitomi Collaboration} {et~al.}(2017){Hitomi Collaboration},
  {Aharonian}, {Akamatsu}, {Akimoto}, {Allen}, {Angelini}, {Audard}, {Awaki},
  {Axelsson}, {Bamba}, {Bautz}, {Blandford}, {Brenneman}, {Brown}, {Bulbul},
  {Cackett}, {Chernyakova}, {Chiao}, {Coppi}, {Costantini}, {de Plaa}, {den
  Herder}, {Done}, {Dotani}, {Ebisawa}, {Eckart}, {Enoto}, {Ezoe}, {Fabian},
  {Ferrigno}, {Foster}, {Fujimoto}, {Fukazawa}, {Furuzawa}, {Galeazzi},
  {Gallo}, {Gandhi}, {Giustini}, {Goldwurm}, {Gu}, {Guainazzi}, {Haba},
  {Hagino}, {Hamaguchi}, {Harrus}, {Hatsukade}, {Hayashi}, {Hayashi},
  {Hayashida}, {Hiraga}, {Hornschemeier}, {Hoshino}, {Hughes}, {Ichinohe},
  {Iizuka}, {Inoue}, {Inoue}, {Ishida}, {Ishikawa}, {Ishisaki}, {Iwai},
  {Kaastra}, {Kallman}, {Kamae}, {Kataoka}, {Katsuda}, {Kawai}, {Kelley},
  {Kilbourne}, {Kitaguchi}, {Kitamoto}, {Kitayama}, {Kohmura}, {Kokubun},
  {Koyama}, {Koyama}, {Kretschmar}, {Krimm}, {Kubota}, {Kunieda}, {Laurent},
  {Lee}, {Leutenegger}, {Limousine}, {Loewenstein}, {Long}, {Lumb}, {Madejski},
  {Maeda}, {Maier}, {Makishima}, {Markevitch}, {Matsumoto}, {Matsushita},
  {McCammon}, {McNamara}, {Mehdipour}, {Miller}, {Miller}, {Mineshige},
  {Mitsuda}, {Mitsuishi}, {Miyazawa}, {Mizuno}, {Mori}, {Mori}, {Mukai},
  {Murakami}, {Mushotzky}, {Nakagawa}, {Nakajima}, {Nakamori}, {Nakashima},
  {Nakazawa}, {Nobukawa}, {Nobukawa}, {Noda}, {Odaka}, {Ohashi}, {Ohno},
  {Okajima}, {Ota}, {Ozaki}, {Paerels}, {Paltani}, {Petre}, {Pinto}, {Porter},
  {Pottschmidt}, {Reynolds}, {Safi-Harb}, {Saito}, {Sakai}, {Sasaki}, {Sato},
  {Sato}, {Sato}, {Sawada}, {Schartel}, {Serlemitsos}, {Seta}, {Shidatsu},
  {Simionescu}, {Smith}, {Soong}, {Stawarz}, {Sugawara}, {Sugita},
  {Szymkowiak}, {Tajima}, {Takahashi}, {Takahashi}, {Takeda}, {Takei},
  {Tamagawa}, {Tamura}, {Tanaka}, {Tanaka}, {Tanaka}, {Tashiro}, {Tawara},
  {Terada}, {Terashima}, {Tombesi}, {Tomida}, {Tsuboi}, {Tsujimoto}, {Tsunemi},
  {Go Tsuru}, {Uchida}, {Uchiyama}, {Uchiyama}, {Ueda}, {Ueda}, {Uno}, {Urry},
  {Ursino}, {de Vries}, {Watanabe}, {Werner}, {Wik}, {Wilkins}, {Williams},
  {Yamada}, {Yamaguchi}, {Yamaoka}, {Yamasaki}, {Yamauchi}, {Yamauchi},
  {Yaqoob}, {Yatsu}, {Yonetoku}, {Zhuravleva}, \& {Zoghbi}}]{Hitomi2017}
{Hitomi Collaboration}, {Aharonian}, F., {Akamatsu}, H., {et~al.} 2017, \nat,
  551, 478

\bibitem[{{Iben} \& {Tutukov}(1984)}]{Iben1984}
{Iben}, I., J. \& {Tutukov}, A.~V. 1984, \apj, 284, 719

\bibitem[{{Iwamoto} {et~al.}(1999){Iwamoto}, {Brachwitz}, {Nomoto},
  {Kishimoto}, {Umeda}, {Hix}, \& {Thielemann}}]{Iwamoto99}
{Iwamoto}, K., {Brachwitz}, F., {Nomoto}, K., {et~al.} 1999, \apjs, 125, 439

\bibitem[{{Janka}(2012)}]{Janka2012}
{Janka}, H.-T. 2012, Annual Review of Nuclear and Particle Science, 62, 407

\bibitem[{{Jordan} {et~al.}(2003){Jordan}, {Gupta}, \& {Meyer}}]{Jordan2003}
{Jordan}, G.~C., {Gupta}, S.~S., \& {Meyer}, B.~S. 2003, \prc, 68, 065801

\bibitem[{{Katz} \& {Dong}(2012)}]{Katz2012}
{Katz}, B. \& {Dong}, S. 2012, arXiv e-prints, arXiv:1211.4584

\bibitem[{{Kirby} {et~al.}(2019){Kirby}, {Xie}, {Guo}, {de los Reyes},
  {Bergemann}, {Kovalev}, {Shen}, {Piro}, \& {McWilliam}}]{Kirby2019}
{Kirby}, E.~N., {Xie}, J.~L., {Guo}, R., {et~al.} 2019, \apj, 881, 45

\bibitem[{{Kobayashi} {et~al.}(2011){Kobayashi}, {Karakas}, \&
  {Umeda}}]{Kobayashi11}
{Kobayashi}, C., {Karakas}, A.~I., \& {Umeda}, H. 2011, \mnras, 414, 3231

\bibitem[{{Kobayashi} {et~al.}(2019){Kobayashi}, {Leung}, \&
  {Nomoto}}]{Kobayashi2019}
{Kobayashi}, C., {Leung}, S.-C., \& {Nomoto}, K. 2019, arXiv e-prints,
  arXiv:1906.09980

\bibitem[{{Kurucz}(2005)}]{Kurucz2005}
{Kurucz}, R.~L. 2005, Memorie della Societa Astronomica Italiana Supplementi,
  8, 14

\bibitem[{{Kushnir} {et~al.}(2013){Kushnir}, {Katz}, {Dong}, {Livne}, \&
  {Fern{\'a}ndez}}]{Kushnir2013}
{Kushnir}, D., {Katz}, B., {Dong}, S., {Livne}, E., \& {Fern{\'a}ndez}, R.
  2013, \apjl, 778, L37

\bibitem[{{Kuuttila} {et~al.}(2019){Kuuttila}, {Gilfanov}, {Seitenzahl},
  {Woods}, \& {Vogt}}]{Kuuttila2019}
{Kuuttila}, J., {Gilfanov}, M., {Seitenzahl}, I.~R., {Woods}, T.~E., \& {Vogt},
  F.~P.~A. 2019, \mnras, 484, 1317

\bibitem[{{Larsen} {et~al.}(2017){Larsen}, {Brodie}, \& {Strader}}]{Larsen2017}
{Larsen}, S.~S., {Brodie}, J.~P., \& {Strader}, J. 2017, \aap, 601, A96

\bibitem[{{Larsen} {et~al.}(2018){Larsen}, {Brodie}, {Wasserman}, \&
  {Strader}}]{Larsen2018}
{Larsen}, S.~S., {Brodie}, J.~P., {Wasserman}, A., \& {Strader}, J. 2018, \aap,
  613, A56

\bibitem[{{Levanon} {et~al.}(2015){Levanon}, {Soker}, \&
  {Garc{\'\i}a-Berro}}]{Levanon2015}
{Levanon}, N., {Soker}, N., \& {Garc{\'\i}a-Berro}, E. 2015, \mnras, 447, 2803

\bibitem[{{Limongi} \& {Chieffi}(2018)}]{LC18}
{Limongi}, M. \& {Chieffi}, A. 2018, \apjs, 237, 13

\bibitem[{{Lind} {et~al.}(2017){Lind}, {Amarsi}, {Asplund}, {Barklem},
  {Bautista}, {Bergemann}, {Collet}, {Kiselman}, {Leenaarts}, \&
  {Pereira}}]{Lind2017}
{Lind}, K., {Amarsi}, A.~M., {Asplund}, M., {et~al.} 2017, \mnras, 468, 4311

\bibitem[{{Lind} {et~al.}(2012){Lind}, {Bergemann}, \& {Asplund}}]{Lind2012}
{Lind}, K., {Bergemann}, M., \& {Asplund}, M. 2012, \mnras, 427, 50

\bibitem[{{Livne}(1990)}]{Livne1990}
{Livne}, E. 1990, \apjl, 354, L53

\bibitem[{{Maoz} {et~al.}(2014){Maoz}, {Mannucci}, \& {Nelemans}}]{Maoz2014}
{Maoz}, D., {Mannucci}, F., \& {Nelemans}, G. 2014, \araa, 52, 107

\bibitem[{{McWilliam}(2016)}]{McWilliam2016}
{McWilliam}, A. 2016, \pasa, 33, e040

\bibitem[{{McWilliam} {et~al.}(2018){McWilliam}, {Piro}, {Badenes}, \&
  {Bravo}}]{McWilliam2018}
{McWilliam}, A., {Piro}, A.~L., {Badenes}, C., \& {Bravo}, E. 2018, \apj, 857,
  97

\bibitem[{{Mernier} {et~al.}(2016){Mernier}, {de Plaa}, {Pinto}, {Kaastra},
  {Kosec}, {Zhang}, {Mao}, {Werner}, {Pols}, \& {Vink}}]{Mernier2016}
{Mernier}, F., {de Plaa}, J., {Pinto}, C., {et~al.} 2016, \aap, 595, A126

\bibitem[{{Mishenina} {et~al.}(2015){Mishenina}, {Gorbaneva}, {Pignatari},
  {Thielemann}, \& {Korotin}}]{Mishenina2015}
{Mishenina}, T., {Gorbaneva}, T., {Pignatari}, M., {Thielemann}, F.-K., \&
  {Korotin}, S.~A. 2015, \mnras, 454, 1585

\bibitem[{{Nomoto} {et~al.}(2013){Nomoto}, {Kobayashi}, \&
  {Tominaga}}]{Nomoto2013}
{Nomoto}, K., {Kobayashi}, C., \& {Tominaga}, N. 2013, \araa, 51, 457

\bibitem[{Oliphant(2007)}]{Oliphant2007}
Oliphant, T.~E. 2007, Computing in Science {\&} Engineering, 9, 10

\bibitem[{{Pakmor} {et~al.}(2012){Pakmor}, {Kromer}, {Taubenberger}, {Sim},
  {R{\"o}pke}, \& {Hillebrandt}}]{Pakmor2012}
{Pakmor}, R., {Kromer}, M., {Taubenberger}, S., {et~al.} 2012, \apjl, 747, L10

\bibitem[{{Pritzl} {et~al.}(2005){Pritzl}, {Venn}, \& {Irwin}}]{Pritzl2005}
{Pritzl}, B.~J., {Venn}, K.~A., \& {Irwin}, M. 2005, \aj, 130, 2140

\bibitem[{{Rebassa-Mansergas} {et~al.}(2019){Rebassa-Mansergas}, {Toonen},
  {Korol}, \& {Torres}}]{Rebassa2019}
{Rebassa-Mansergas}, A., {Toonen}, S., {Korol}, V., \& {Torres}, S. 2019,
  \mnras, 482, 3656

\bibitem[{{Romano} {et~al.}(2010){Romano}, {Karakas}, {Tosi}, \&
  {Matteucci}}]{Romano10}
{Romano}, D., {Karakas}, A.~I., {Tosi}, M., \& {Matteucci}, F. 2010, \aap, 522,
  A32

\bibitem[{{Ruiter} {et~al.}(2009){Ruiter}, {Belczynski}, \&
  {Fryer}}]{Ruiter2009}
{Ruiter}, A.~J., {Belczynski}, K., \& {Fryer}, C. 2009, \apj, 699, 2026

\bibitem[{{Sbordone} {et~al.}(2004){Sbordone}, {Bonifacio}, {Castelli}, \&
  {Kurucz}}]{Sbordone2004}
{Sbordone}, L., {Bonifacio}, P., {Castelli}, F., \& {Kurucz}, R.~L. 2004,
  Memorie della Societa Astronomica Italiana Supplementi, 5, 93

\bibitem[{{Scalzo} {et~al.}(2019){Scalzo}, {Parent}, {Burns}, {Childress},
  {Tucker}, {Brown}, {Contreras}, {Hsiao}, {Krisciunas}, {Morrell}, {Phillips},
  {Piro}, {Stritzinger}, \& {Suntzeff}}]{Scalzo2019}
{Scalzo}, R.~A., {Parent}, E., {Burns}, C., {et~al.} 2019, \mnras, 483, 628

\bibitem[{{Scalzo} {et~al.}(2014){Scalzo}, {Ruiter}, \& {Sim}}]{Scalzo2014}
{Scalzo}, R.~A., {Ruiter}, A.~J., \& {Sim}, S.~A. 2014, \mnras, 445, 2535

\bibitem[{{Seitenzahl} {et~al.}(2013){Seitenzahl}, {Cescutti}, {R{\"o}pke},
  {Ruiter}, \& {Pakmor}}]{Seitenzahl2013}
{Seitenzahl}, I.~R., {Cescutti}, G., {R{\"o}pke}, F.~K., {Ruiter}, A.~J., \&
  {Pakmor}, R. 2013, \aap, 559, L5

\bibitem[{{Seitenzahl} \& {Townsley}(2017)}]{Seitenzahl2017}
{Seitenzahl}, I.~R. \& {Townsley}, D.~M. 2017, {Nucleosynthesis in
  Thermonuclear Supernovae}, ed. A.~W. {Alsabti} \& P.~{Murdin}, 1955

\bibitem[{{Shappee} {et~al.}(2013){Shappee}, {Kochanek}, \&
  {Stanek}}]{Shappee2013}
{Shappee}, B.~J., {Kochanek}, C.~S., \& {Stanek}, K.~Z. 2013, \apj, 765, 150

\bibitem[{{Shen} {et~al.}(2018){Shen}, {Kasen}, {Miles}, \&
  {Townsley}}]{Shen2018}
{Shen}, K.~J., {Kasen}, D., {Miles}, B.~J., \& {Townsley}, D.~M. 2018, \apj,
  854, 52

\bibitem[{{Sneden} {et~al.}(2016){Sneden}, {Cowan}, {Kobayashi}, {Pignatari},
  {Lawler}, {Den Hartog}, \& {Wood}}]{Sneden2016}
{Sneden}, C., {Cowan}, J.~J., {Kobayashi}, C., {et~al.} 2016, \apj, 817, 53

\bibitem[{{Spite} {et~al.}(2013){Spite}, {Caffau}, {Bonifacio}, {Spite},
  {Ludwig}, {Plez}, \& {Christlieb}}]{Spite2013}
{Spite}, M., {Caffau}, E., {Bonifacio}, P., {et~al.} 2013, \aap, 552, A107

\bibitem[{{Takeda} {et~al.}(2005){Takeda}, {Hashimoto}, {Taguchi}, {Yoshioka},
  {Takada-Hidai}, {Saito}, \& {Honda}}]{Takeda2005}
{Takeda}, Y., {Hashimoto}, O., {Taguchi}, H., {et~al.} 2005, \pasj, 57, 751

\bibitem[{{Thielemann} {et~al.}(1986){Thielemann}, {Nomoto}, \&
  {Yokoi}}]{Thielemann1986}
{Thielemann}, F.-K., {Nomoto}, K., \& {Yokoi}, K. 1986, \aap, 158, 17

\bibitem[{{Toonen} {et~al.}(2018){Toonen}, {Perets}, \& {Hamers}}]{toonen2018}
{Toonen}, S., {Perets}, H.~B., \& {Hamers}, A.~S. 2018, \aap, 610, A22

\bibitem[{{Travaglio} {et~al.}(2004){Travaglio}, {Hillebrandt}, {Reinecke}, \&
  {Thielemann}}]{Travaglio2004}
{Travaglio}, C., {Hillebrandt}, W., {Reinecke}, M., \& {Thielemann}, F.-K.
  2004, \aap, 425, 1029

\bibitem[{{Whelan} \& {Iben}(1973)}]{Whelan1973}
{Whelan}, J. \& {Iben}, Icko, J. 1973, \apj, 186, 1007

\bibitem[{{Woosley} \& {Weaver}(1995)}]{WW95}
{Woosley}, S.~E. \& {Weaver}, T.~A. 1995, \apjs, 101, 181

\end{thebibliography}

\end{document}